\documentstyle[preprint,pre,aps]{revtex}
\begin{document}
\draft
\title{Non-equilibrium phase transitions\\
in one-dimensional kinetic Ising models}
\author{N.Menyh\'ard}
\address{Research Institute for Solid State Physics, \\
H-1525 Budapest, P.O.Box 49, Hungary}
\author{G. \'Odor}
\address{Research Institute for Materials Science, \\
H-1525 Budapest, P.O.Box 49, Hungary}
\maketitle
\begin{abstract}
A family of nonequilibrium kinetic Ising models, introduced earlier,
evolving under the
competing effect
of spin flips at {\it zero temperature} and nearest
neighbour random spin exchanges
 is further investigated here.
By increasing the range of spin exchanges and/or their strength
the nature of the
phase transition 'Ising-to-active' becomes of (dynamic) mean-field
type and a  first order tricitical point is located at the Glauber
($\delta=0$) limit.
 Corrections
to mean-field theory are evaluated up to sixth order
in a cluster approximation and found to give
good results concerning the phase boundary and the critical
exponent $\beta$ of the order parameter which is obtained
as $\beta\simeq1.0$.

\end{abstract}
\narrowtext
\section{Introduction}

Kinetic Ising models were originally intended to study relaxational
processes near equilibrium states \cite{gla63,kaw72}. Later combinations
of Glauber and Kawasaki dynamics were used successfully in investigating
questions about temperature driven nonequilibrium phase transitions
\cite{dem85,gon87,wan88}.
In a previous paper
 a class of general nonequilibrium kinetic Ising models
(NEKIM) with combined spin flip dynamics at $T=0$ and spin
exchange dynamics at $T=\infty$ has been introduced \cite{men94}, in which,
for a range of parameters ( other then temperature)
of the model, a directed-percolation-like  Ising-to-active
phase transition takes place.
The line of phase transitions have been found to belong to the
same universality class as the phase transitions  occurring in the
cellular automaton
models introduced and investigated by Grassberger et al.
\cite{gra84,gra89}.
 Numerical studies of  other models showing
similar type of phase transition have been reported
recently \cite{jen94,kim94}.

In the present note we consider a generalized form of
NEKIM by allowing for exchanges of arbitrary range, $R$.
The mean-field (MF) limit of NEKIM phase transitions is reached when
$R\rightarrow\infty$ and/or the probability of the exchanges, $p_{ex}$,
relative to the time scale  of spin-flips
approaches infinity.
 In a systematic generalized MF approach (GMF)\cite{gut87,dic88,sza91},
 besides
the lowest order approximation (ordinary dynamic MF, $n=1$)
also the second order cluster equations(n=2)  could be solved exactly.
Numerical solutions have been obtained up to sixth order.

In this way we have found strong theoretical evidence for the
conjecture, stemming from simulations \cite{men94},
 that  the line of Ising-to-active second
order phase transition points ends at
 the Glauber limit ($\delta_{c}=0$,
$\delta$ being a parameter of the spin-flip transition rate of crucial
importance here)
with maximal exchange range and/or rate. It is shown here that this end point
is of first order (tricritical point) and is
described by plain MF theory. The
relaxation time is obtained as $\tau\propto 1/{\mid\delta\mid}$.
 GMF results show, that with increasing $n$
the critical point becomes of second order and
moves towards  negative values of
$\delta$ of increasing absolute value.
The coherent anomaly method \cite{suz86,kol94} has been used to extract
the exponent $\beta$ of the order parameter
 from the results of GMF calculations
yielding  $\beta=1.0$.

\section{The model}
\label{sec:2}
In NEKIM we have started with the general form of the Glauber \cite{gla63}
spin-flip transition rate
in one-dimension
for spin $s_i$
sitting at site $i$  ($s_i=\pm1$):
\begin{equation}
w_i = {{\Gamma/2}}(1+\delta s_{i-1}s_{i+1})\left(1 -
{\gamma\over2}s_i(s_{i-1} + s_{i+1})\right)
\end{equation}
where $\gamma=\tanh{{2J}/{kT}}$ ($J$ denoting the coupling constant in
the Ising Hamiltonian), $\Gamma$ and $\delta$ are further
 parameters which can, in general,
 also depend on temperature. When $T=0$ is taken then
$\gamma=1$ and (1)
leads to two independent rates:
\begin{equation}
p_{RW}\equiv{2w_{\uparrow\downarrow\downarrow}}={\Gamma}(1-\delta),\,\,
 p_{an}\equiv{w_{\uparrow\downarrow\uparrow}}={\Gamma}(1+\delta)
 \end{equation}
  responsible for random walk and
pairwise annihilation of kinks, respectively.
$\Gamma$ and $\delta$ are constants to be varied.\\
The other ingredient of NEKIM has been
a spin-exchange transition rate of neighbouring spins (
the Kawasaki\cite{kaw72} rate at $T=\infty$):
\begin{equation}
w_{ii+1}={1\over2}p_{ex}[1-s_is_{i+1}]
\end{equation}
where $p_{ex}$ is the probability of spin exchange.
 $p_{RW}$, $p_{an}$ and $p_{ex}$ have
been chosen
as normalised to unity, leading to the relation:
\begin{equation}
p_{ex}=1-2\Gamma
\end{equation}
 The spin-exchange process
induces
 pairwise creation of kinks in the immediate neighbourhood of an
existing kink: $k \rightarrow 3k$ with probability
${p_{ex}}$. From this process the ultimate development
of an active phase can arise and
 in ref \cite{men94} we have made the conjecture,
and found numerical evidence for it,
that $p_{RW}>p_{an}$ (i.e. $\delta<0$) is
 necessary for this to happen.

Now we  generalize the original NEKIM model by
 allowing the range of the
spin-exchange to vary. Namely, eq.(3) is replaced by
\begin{equation}
w_{i,i+k}={1\over2}p_{ex}[1-s_{i}s_{i+k}],
\end{equation}
 where $i$ is a randomly chosen
site and $s_i$ is allowed to exchange with $s_{i+k}$ with
probability $p_{ex}$. Site $k$ is again randomly chosen in the
interval $1\leq R$ , $R$ being thus the range of exchange.
The spin-flip part of the model will be  unchanged.
We have carried out numerical studies with this generalized
model in order to locate the lines of Ising-to-active phase
transitions.
 Spin-flip and spin-exchange have been applied
alternatingly at each time step, the spin-flip part has been
applied using two-sublattice updating, while making $l$
 Monte Carlo attempts at random ( $l$
denotes the size of the chain) has been counted as one time-step
of exchange updating.
It is worth
mentioning, that besides $k\rightarrow3k$, also
the process $k\rightarrow5k$ can occur for $R\geq3$, and
the new kink pairs are not necessarily neighbors. The character
of the phase transition line at $R>1$ is similar to that for $R=1$,
except that
the active phase extends, asymptotically, down to $\delta_c=0$.
This is illustrated in Fig.1 , where besides $R=1$, the
case $R=3$ is also depicted: the critical value of $-\delta_{c}$ is
shown as a function of $p_{ex}$  (Fig 1.a) and
b)). Moreover, $-\delta_{c}$ as a function of $R$ is also
shown at constant $\Gamma=.35$ . The abscissa has been suitably
chosen
to squeeze the whole ( infinite ) range of $R$ between $0$ and $1$
and
 for getting phase lines of comparable size (hence the power 4 of
$R/(1+R)$ in case of Fig.1.c)).

Besides the critical exponent $\alpha$, used in identifying the
phase transition points, also the other
 critical exponents characterizing the phase transition
 have  been determined numerically around some  typical points
(far from the end-points) of the phase transition lines for $R>1$,
with the same result as
was obtained  in \cite {men94} for the case $R=1$: the exponents
 agree  - within error -  with those of Grassberger's automata
\cite{gra84,gra89}.

On the phase diagrams of Fig.1 two non-typical regimes can be
distinguished, namely\\
a). $p_{ex}\approx 0$.( on Figs. 1.a) and b).)
 Here NEKIM's behaviour is getting close to
that of the plain spin-flip
model at $T=0$: the steady state is everywhere Ising-like except for
the limit-point $\delta=-1$, where $p_{an}=0$
and the initial kink
density is sustained. At this specific point the energy
becomes conserved \cite{spo89} and a phase transition takes place
with a change in the form of the time dependence of correlations
from exponential to stretched exponential.
This limit will not be further  discussed here.

b). $p_{ex}\approx 1.0$ and/or $R\rightarrow\infty$ (on Figs 1.a)-c)).
 For $p_{ex}\rightarrow1$, $R=1$ we have concluded in [6] that
  $\delta_{c}\rightarrow{-0}$, though it has been difficult
 to get reliable numerical estimates
for the critical exponents of the transition due to the long transients.

In  limit b)., after each step of
spin-flip ordering, maximal mixing of the neighbourhood of
each spin follows, suggesting that a mean-field type situation
takes place. It is important that according to eq.(4) $p_{ex}=1$
is approached together with $\Gamma\rightarrow0$ and thus
 $p_{ex}/{\Gamma}\rightarrow
\infty$. (As $\Gamma$ sets the time scale of
the ordering process, its vanishing tendency enhances the effect of
mixing).
The same limit can be reached at fixed $p_{ex}$ by increasing $R$ to
infinity (Fig.1.c)).

  We have also checked the decrease of
$-\delta_c$ with $1/R$
numerically at fixed $\Gamma=.35, p_{ex}=.3$  and found,
over the decade of $R=4 - 40$, that
\begin {equation}
{-\delta_c}\approx{2.0{(1/R)}^2}
\end{equation}
reminiscent of a crossover type behaviour of equilibrium and non-equilibrium
phase transitions \cite{rac94}, here with crossover exponent $1/2$.
It should be
noted here, that to get closer to the expected $\delta_{cMF}=0$, longer chains
(we used $l$ values up to $20000$) and longer runs (here up to $t=5\cdot10^4$)
would have been necessary.
 The former to ensure $l\gg R$ \cite{mon93}
and the latter to overcome the long transients
present at the first few decades of time steps.

 In what follows we will always refer to the MF limit in
connection with $p_{ex}\rightarrow 1$ (i.e. $p_{ex}/{\Gamma}\rightarrow
\infty$), for the sake of
concreteness, but keep in mind that $R\rightarrow \infty$
can play the same role.

\section{Mean-field theory and corrections to mean-field}
\label{3}
It is straightforward to find the MF equation for the
spin-flip model alone ( at $T=0$).
By denoting the average spin density by $M$ we get, using (1)
\begin{equation}
{dM/{dt}}=-{\delta\Gamma}M(M^2-1)
\end{equation}
The fixed point solutions are $M^{*}=1,-1,0$ of which the first
two are stable if $\delta>0$, while the $M^{*}=0$ solution is stable for
$\delta<0$, suggesting a (first order) order-disorder-type phase
transition at $\delta=0$. That the  $M^{*}=0$ fixed point is not an
antiferromagnetic
type can be shown by carrying out a two-sublattice MF analysis
of the model [19].
First sublattice: odd lattice points with average magnetization
$M_1$, second sublattice: even lattice points with average magnetization
$M_2$. The total average magnetization $M={(M_1+M_2)}/2$ and the difference
of the sublattice
magnetizations $\Delta={(M_1-M_2)}/2$ obey the following MF equations:
\begin{equation}
{dM/{dt}}=-{\delta{\Gamma}}M(M^2-1-{\Delta}^2)
\end{equation}
\begin{equation}
{d\Delta/{dt}}=-{2\Gamma\Delta}+{\delta{\Gamma}}\Delta{(M^2-1-
{\Delta}^2)}
\end{equation}
The solutions for fixed point $\Delta^{*}\neq0$
are:  $M^{*}=0$, ${\Delta^{*}}^2=-1-{2\over\delta}$.

The values of $\delta$ being
restricted to $1\geq\delta\geq {-1}$, the only possibility
to ensure ${\Delta^{*}}^2>0$ at the same time is: $ \delta=-1$,
with ${\Delta^{*}}^2=1$. Thus we will suppose that the  transition
at $\delta_c=0$
is
of order-disorder type.
A small fluctuation $dM$ around one of the stable fixed points decreases
as ${dM}{\propto}{e^{-t/\tau}}$ with $\tau=1/{\Gamma{\mid\delta\mid}}$.
This relaxation time becomes infinite at the MF transition point.
The corresponding critical slowing down in its vicinity explains
the longer and longer transients observed during simulations.

Fig. 2. serves to illustrate the MF result in comparison with
results of simulation of NEKIM at three values of $p_{ex}$.
The average density of kinks at $t=\infty$  is depicted
versus $\delta$. MF approximation corresponds to Fig. 2.a). with a jump at
$\delta_{cMF}=0$. Fig. 2.b). shows the behaviour of
the pure spin-flip model at $T=0$: the steady state is everywhere
Ising-like ($\rho(\infty)=0$) except for $\delta=-1$.  Figs. 2.c).,d). and e.)
are results of simulation of NEKIM at $p_{ex} =.02$ ,
$p_{ex}=.9$ and
$p_{ex}=.98$,  respectively. By further
decreasing (increasing) $p_{ex}$, the NEKIM curves  get closer and
closer to b). ( a).), respectively.
 Such behaviour is
in accordance with our
expectations: it   supports
 the MF interpretation of the high-$p_{ex}$ part of the phase
diagram.

We have applied  the generalized
mean-field calculation method, or cluster approximation \cite{gut87,dic88}
in the form applied for cellular automata \cite{sza91}
in order to go beyond the
lowest order approximation shown above.

Steady-state equations have been set up for block probabilities
in $n=2...6$-th order. The system of GMF equations are solvable
analitically also for $n = 2$.

The $n = 2$ approximation gives the density of kinks $\rho(\infty)$ as :
\begin {equation}
\rho(\infty) ={{{3\over 4}\,{{{ p_{RW}}}^2} + { p_{an}} - { p_{RW}}\,{ p_{an}}
-
     {\sqrt{{{{{1\over 16} p_{RW}}}^4} + {3\over 2}\,{{{ p_{RW}}}^2}\,{ p_{an}}
-
         {1\over 2}\,{{{ p_{RW}}}^3}\,{ p_{an}} + {{{ p_{an}}}^2} -
         2\,{ p_{RW}}\,{{{ p_{an}}}^2}}}}\over
   {2\,\left({1\over 2}\,{{{ p_{RW}}}^2} - { p_{RW}}\,{ p_{an}} + {{{
p_{an}}}^2}
        \right) }}
\end{equation}
for $\delta<0$. For $\delta>0$ $\rho(\infty)=0$, i.e. GMF  still
predicts a first order transition for $\delta = 0$; the
jump in $\rho(\infty)$ at $\delta=0$, however, decreases
  monotonically with decreasing $\Gamma$, according to eqs. (10) and (2).

In order to get the $n > 2$ approximations, the set of GMF
 equations can be solved
numerically only. We determined the solutions of the $n =3,4,5,6$
 approximations for the kink density at i).$\Gamma = 0.35$ (Figure 3.) and
of the $n=3,4,5$ approximations at ii).$\Gamma=.05$ (Figure 4.). As
 we can see the transition curves became continuous, with
negative values for $\delta_{c}^n$ ($\delta_{c}^n$ denotes
the value of $\delta$ in the $n$-th approximation for which
the corresponding $\rho(\infty)$ becomes zero).
 Moreover, $\mid\delta_{c}^n\mid$
increases  with growing $n$ values.  As increasing $n$
corresponds to decreasing mixing, i. e. decreasing $p_{ex}$,
the tendency shown by the above results is correct.

Figs. 5.a). and 5.b). show
a quantitative - though only tentative -
comparison between the results of GMF and the simulated NEKIM
phase diagrams. The  obtained GMF data for $\delta_{c}^n$ corresponding
 to $n=3,4,...,6$ ($\Gamma=.35$) are
depicted in Fig. 5.a.) as a function of $1/(n-3)$, together with results of
simulations.
 The correspondence between $n$ and $p_{ex}$ has
been chosen as the simplest conceivable one. ( Note that $\delta_c\not=0$
is obtained first for $n=4$.). The simulated phase diagram has been
obtained without requiring the fulfillment of  eq.(4), at constant
 $\Gamma=.35$. In this
case the $\delta_c=0$ limit, of course, is not reached and a purely second
order phase transition line can be compared with GMF results
(for $n$ values where it also predicts a second order transition).
 Simulations for $R=3$ have been found to lead to $\mid\delta_c\mid$
values low enough
to fit GMF data. The (polynomial) extrapolation of GMF data  to
$n\rightarrow\infty$  (corresponding to $p_{ex}=0$ , i.e. plain spin-flip),
shown also in Fig. 5., could have been expected to approach $\delta=-1$.
That this is not case
 can be due to the circumstances that upon
increasing $n$ i)
GMF starts here from a first-order MF phase transition,
 ii). which becomes second
order and iii).GMF  should end up at a pathological point, discussed shortly
in section II., with quite unusual (and not yet cleared up) properties.

Fig 5b). shows the $n=3,4,5$ results
for $\Gamma=.05$, which are compared now with
the $R=1$ simulation data.

The critical exponent  $\beta$ of the order parameter
 has been determined processing
the results of GMF approximation by the Coherent-Anomaly
Method (CAM) [14,15].
According to CAM the GMF solution for kink density $\rho$ at a
given level of approximation -- in the vicinity of the critical
point, $\delta_c$ -- is the product of some mean-field
 behavior multiplied by the anomaly factor $a(n)$:
\begin{equation}
\rho(n) = a(n) \ (\delta / \delta_c^n - 1)^{\beta_{MF}} \ ,
\end{equation}
The true critical exponent, $\beta$, can be obtained by
fitting, using the knowledge that the divergence of the anomaly
factor scales as :
\begin{equation}
a(n) \sim (\delta_c^n / \delta_c - 1)^{\beta - \beta_{MF}} \ ,
\label{ano}
\end{equation}
as the level of approximation $n$ goest to infinity.
More precisely, for the available low level of approximations ($n \le 6$),
 correction to scaling should also be taken into account :
\begin{equation}
a(n) = b \ \Delta_n^{\beta - \beta_{MF}} + c \ \Delta_n^{\beta - \beta_{MF}
+ 1} + ... \ ,
\end{equation}
where $b$ and $c$ are constants and the invariant variable
\begin{equation}
\Delta_n = (\delta_c / \delta_c^n)^{1/2} - (\delta_c^n / \delta_c)^{1/2}
\end{equation}
is used. This new variable was introduced recently \cite{kol94} to avoid
the ambiguity on the choice of the independent variable ($\delta
\leftrightarrow \delta^{-1}$). Using this new variable accurate estimate
was given for the critical exponents of the 3D Ising model \cite{kol94}
and for the exponent $\beta$ of the stochastic
Rule 18 cellular automaton \cite{odo95}.

 From our GMF approximation results , as shown on Fig.3, we can use
the $n=4,5,6$ data
for the CAM analysis, while the $n=3$ result can be taken
to represent the lowest order MF approximation
(with ${\delta_c}^{MF}=0$) for a
{\it continuous} transition ( no jump in $\rho$ for $n=3$).
For $\delta_c$ we use the results of the polynomial extrapolation.
Fig.6. shows that in the $n=3$ approximation the exponent $\beta=1.0064$
, thus $\beta_{MF}\approx1$.
Graphs similar to that on Fig.6.
 have been obtained for $n=4,5,6$, as well.  Consequently,
as Table \ref{tablex} shows, the anomaly factor
does not depend on $n$. This means, according to eq.(12), that
the exponent is estimated to be equal to the 'mean-field' value
$\beta \simeq \beta_{MF} = 1$.
\nopagebreak
\narrowtext
\begin{table}
\caption{CAM calculation results }
\begin{tabular}{lrl}
$n$ & $\Delta_c^n$   & $a(n)$ \\
\tableline
$4$ & $2.49043$  & $0.01083$  \\
$5$ & $1.81022$  & $0.01074$  \\
$6$ & $1.45766$  & $0.01079$  \\
\end{tabular}
\label{tablex}
\end{table}
\pagebreak
\section{Discussion}
\label{4}
The mean-field limit of the line of non-thermal phase transitions
occurring in a
family of one-dimensional kinetic Ising models has been analysed here.
This line consists of second order Ising-to-active phase transition
points which belong to the  universality class found first by
Grassberger et al\cite{gra84,gra89}.
The first order endpoint of this line has been found to be described by
MF theory.  Systematic generalized MF theory has been applied
to treat bigger and bigger blocks of size $n$  exactly  in order to
be able to depart from this tricritical point.
Numerically solvable results up to $n=6$ have given support of
results of simulations and especially provided a value
for the critical exponent $\beta$ of the order parameter (density
of kinks) $\beta=1$, which is in accord with Grassberger's result :
$\beta=.94\pm .06$. The value $\beta=1$   coincides with the MF
 $\beta$-exponent for directed percolation. This is not surprising
in case of our $n=3$ result which we have used  as an effective
MF one for a {\it continuous} transition at $\delta=0$. As Grassberger has
pointed out \cite{gra89} in the rate (or MF) approximation
there is no difference between models leading to the Ising-to-
active transition and directed percolation. In this argumentation,
however, the MF equation is written for the kink density (and not for
the magnetization as in eq.(7)) and has  the form:
 $ dn/dt= 2\mu n-2\lambda n^2 $, where $\mu$ and $\lambda$
are the reproduction and annihilation rates, respectively.
Nevertheless, a heuristic equation of similar type can also be constructed
in the present model
using $\mu\propto (p_{RW}-p_{an})$ as a rate making kink-reproduction
effective (a conclusion stemming from simulations). The corresponding
Mf critical value is then $\delta_{c}^{MF}=0$ and $\beta^{MF}=1$.
It is, however, surprising that higher order approximations of GMF
have not given practically any deviation from the MF value of $\beta$.
   In simulations for branching annihilating random walk
with four offsprings Jensen \cite{jen94} conjectures a value
$\beta=13/14$ on the basis of simulation results, no theoretical
motivation appears to exist for this value, however. To decide
the question what the exact value of $\beta$ in this universality
class is appears to be a challenging task, and probably  calculating
GMF in even higher approximations than here would be worth wile.

\acknowledgments
The authors would like to thank the Hungarian research fund OTKA ( Nos.
T-2090, T-4012 and F-7240) for
support during this study. One of us (N.M.) would like to acknowledge
support by Deutsche Forschunggemeinschaft, SFB341, during her
stay in K\"oln, where part of this work has been carried out.

\begin{figure}
\caption{  Phase diagram of NEKIM for $\delta_c{(R,p_{ex})}$ is
depicted for
a). $R=1$, b). $R=3$ as a function of $p_{ex}$ (with $2\Gamma=1-p_{ex}$)
and for c). $\Gamma=.35$ ($p_{ex}=.3$)
as a function of ${(R/{1+R})}^4$ (full line).
 The phase
boundaries have been obtained by measuring $\rho(t)$, the density of
kinks, starting from a random initial distribution and locating the
phase transition points by $\rho(t)\propto t^{-\alpha}$ with
$\alpha=.27\pm.04$. Typically the  number of lattice points has been
$l=2000$  and averaging over 500  independent runs
has been performed.}\label{fig.1}
\end{figure}
\begin{figure}
\caption { Kink density $\rho$ in the steady state as a function of
$\delta$: a).result of first order MF calculation with a jump
at $\delta=0$, reflecting the first order nature of the transition,
b). the same quantity for the pure spin-flip model: $\rho(\infty)=
\rho(0)$ at $\delta=-1$, otherwise $\rho(\infty)=0$.
c).,d).,e).: results of simulations of NEKIM, see text.}\label{fig.2}
\end{figure}
\begin{figure}
\caption{ $\rho(\infty)$ as a function of $\delta$ as obtained from
GMF for $n=3,..,6$ in case of $\Gamma=.35$}\label{fig.3}
\end{figure}
\begin{figure}
\caption{ Kink density $\rho(\infty)$ obtained by GMF for
$n=3,4,5$ in case of $1\Gamma=.05$}\label{fig.4}
\end{figure}
\begin{figure}
\caption { Comparison between results of simulation and GMF:
GMF results for $\delta_c$ are plotted as a function of $1/{(n-3)}$,
while simulation results for $\delta_c$ are depicted as a function
of $p_{ex}$
 at {\it constant}
$\Gamma$, with $R=3$ in case of $\Gamma=.35$ (Fig. 5.a)) and with
$R=1$ for $\Gamma=.05$ (Fig. 5.b).).}
\label{fig.5}
\end{figure}
\begin{figure}
\caption{The $n=3$ results of  the GMF for  $\rho(\infty)$ with
 $\Gamma=.35$, are shown on a double logarithmic plot
as a function of $\mid(\delta-\delta_c)\mid$. The straight line
  gives $\beta=1.0064$, which is
used as a fictitious MF value in CAM.}\label{fig.6}
\end{figure}

\begin{references}
\bibitem{gla63}    Glauber R J  J.Math.Phys.{\bf 4}
 (1963) 191
\bibitem{kaw72}  see e.g. {\it Kawasaki K:}
 Phase Transitions and Critical Phenomena,Vol.2.,\hfil
 ed.{\it Domb C and Green M S}
 (New York: Academic, 1972) p.443
\bibitem{dem85} DeMasi A.,  Ferrari P.A. and  Lebowitz J.L.,
Phys.Rev.Lett. {\bf 55} (1985) 1947; J.Stat.Phys. {\bf 44} (1986) 589.
\bibitem{gon87} Gonzalez-Miranda J.M., Garrido P.L., Marro J. and
Lebowitz J.L., Phys.Rev.Lett. {\bf 59} (1987) 1934.
\bibitem{wan88} Wang J.S. and Lebowitz J.L., J.Stat.Phys. {\bf 51}
(1988) 893.
\bibitem{men94}  Menyh\'ard N., J.Phys.A:Math.Gen.{\bf 27} (1994) 6139
\bibitem{gra84}  Grassberger P., Krause F. and von der Twer T.\\
J.Phys.A:Math.Gen.{\bf 17}(1984)L105
\bibitem{gra89}  Grassberger P.,  J.Phys. A:Math.Gen.
{\bf 22} (1989) L1103
 \bibitem{jen94} Jensen I., University of Melbourne Preprint, May 1994
\bibitem{kim94} Kim M.H., Park H.,Critical Behaviour of an Interacting
Monomer-Dimer Model, cond-mat/9312048
\bibitem{gut87}Gutowitz H.A., Victor J.D., Knight B.W. ,Physica {\bf
28D}, (1987), 18
\bibitem{dic88}Dickman R., Phys.Rev. {\bf A38} (1988) 2588
\bibitem{sza91}Szab\'o G., Szolnoki A. and Bod\'ocs L. Phys.Rev.
{\bf A 44} (1991) 6375
Szab\'o G., \'Odor G. Phys.Rev. {\bf E59} (1994) 2764 and references
therein
\bibitem{suz86}  Suzuki M., J.\ Phys.\ Soc. Jpn.\ {\bf 55}, 4205 (1986).
\bibitem{kol94} Kolesik M. and  Suzuki M., cond-mat/9411109.
\bibitem{spo89} Spohn H. Commun. Math. Phys. {\bf125} (1989) 3
\bibitem {rac94}Racz Z. and  Zia K.P., Phys.Rev.E {\bf 49} (1994) 139
\bibitem{mon93} Mon K.K. and Binder K., Phys.Rev. {\bf 48E} (1993) 2498
\bibitem {men88} Menyh\'ard N. in Proceedings of the Conference
on Synergetics, Order and Chaos, Madrid, 1987 ed.:
M.G.Velarde, World Scientific Publ., Singapore, 1988, pp. 590-600.
\bibitem{odo95}G. \'Odor G. to appear in  Phys.\ Rev.\ E.

\end{references}
\end{document}